\documentclass[aps,preprint]{revtex4}
\usepackage{amsfonts}
\usepackage{amssymb,epsfig}
\usepackage{latexsym}
\usepackage{amsmath}
\usepackage{graphicx}

\begin{document}

\title{Agegraphic Reconstruction of Modified $F(R)$ and $F(\mathcal{G})$ Gravities}
\author{A. Khodam-Mohammadi\footnote{E-mail: \texttt{khodam@basu.ac.ir}},~~P. Majari~~and\
 M. Malekjani\footnote{E-mail: \texttt{malekjani@basu.ac.ir}}}
\address{Physics Department, Faculty of Science, Bu-Ali Sina
University, Hamedan , Iran}

\begin{abstract}
The cosmological reconstruction of modified $F(R)$ and
$F(\mathcal{G})$ gravities with agegraphic dark energy (ADE) model
in a spatially flat universe without matter field is investigated by
using e-folding "$N$" as a forward way. After calculating a
consistent $F(R)$ in ADE's framework, we obtain conditions for
effective equation of state parameter $w_{\mathrm{eff}}$, and see
that reconstruction is possible for both phantom and non-phantom
era. These calculations also are done for $F(\mathcal{G})$ gravity
and the condition for a consistent reconstruction is obtained.
\end{abstract}


\maketitle

\section{introduction}
Dark energy problem attracted a great deal of attention at the last
decade. Recent astrophysical data suggest that our universe behave
under an accelerated expansion with an effective equation of state
parameter $-1.48<w_{\mathrm{eff}}<-0.72$ \cite{Hann}. Up to now,
scientists have proposed two prescriptions for this expansion. One
group believe that a component of dark energy (DE), which possesses
negative pressure, is the source of this expansion. It is understood
that about 70 percent of energy content of the current universe is
dark energy. Several models of dynamical dark energy, after
$\Lambda$-CDM model, whose equations of states are no longer a
constant but evolve with time, have been proposed by this group. On
the other hand, a curvature driven acceleration model which is
called, modified gravity, has been proposed  by Strobinsky
\cite{Stra} and Kerner \cite{Ker} et al., for the first time, in
1980. Modified gravity approach suggests the gravitational
alternative for unified description of inflation, dark energy and
dark matter with no need of the hand insertion of extra dark
components. It has been shown that these two approaches may related
to each other. Many authors have extended a reconstruction technique
in order to made a correspondence between an acceptable cosmological
model, and a modified gravity \cite{Noj2,Cruz,Noj1,Bam}. Also a
reconstruction scheme has been developed in terms of e-folding (or
redshift $z$), and some generalization of such technique for viable
$F(R)$ gravity has been done, so that local tests were usually
satisfied \cite{Noj3}. By using this technique, some of works have
been presented where $F(R)$ and $F(\mathcal{G})$ gravities are
reconstructed so that they give the well-known cosmological
evolution. The $\Lambda$CDM epoch, deceleration/acceleration epoch
which is equivalent to presence of phantom and non-phantom matter,
late-time acceleration with the crossing of phantom-divide line
\cite{Noj3,Eliz} and the holographic dark energy model \cite{Set1}
are some of examples that have been presented in the recent years.
The agegraphic dark energy (ADE) model is one of the interested
model, which has been welcomed by many authors. The cosmological
behavior, statefinder analysis \cite{Khod1} and other cosmological
aspects of ADE model have been calculated in an
interacting/non-interacting spatially flat/non-flat,
ordinery/entropy-corrected versions of Friedman-Robertson-Walker
(FRW) universe \cite{Karami1,Malek2}. ADE model is arisen from
combining quantum mechanics with general relativity, directly. This
model, proposed by Cai \cite{Cai1}, is based on the line of quantum
fluctuations of spacetime, the so-called K\'{a}rolyh\'{a}zy relation
$\delta t=\lambda t_{p}^{2/3}t^{1/3}$, and the energy-time
Heisenberg uncertainty relation $E_{\delta t^{3}}\sim t^{-1}$.
Throughout this paper, we use the Planck unit ($\hbar =c=k_{B}=1$) ,
where $t_{p}=l_{p}=1/m_{p}$ are Planck's time, length and mass,
respectively. These relations enable one to obtain an energy density
of the metric quantum fluctuations of Minkowski spacetime as follows
\cite{Maz}
\begin{equation}
\rho _{q}\sim \frac{E_{\delta t^{3}}}{\delta t^{3}}\sim \frac{1}{%
t_{p}^{2}t^{2}}\sim \frac{m_{p}^{2}}{t^{2}}.  \label{ED}
\end{equation}%
In ADE, this energy density is considered as density of dark energy
component, $\rho _{d}$, of spacetime. By considering a FRW universe,
due to effect of curvature, one should introduce a numerical factor
$3n^{2}$ in (\ref{ED}) \cite{Cai1, Cai3}.

In this paper we want to reconstruct a consistent modified gravity
so that it gives the cosmological evolution of ADE model. Specially
we consider $F(R)$ and modified Gauss-Bonnet (GB) $F(\mathcal{G})$
gravities.
\section{the formalism of modified gravity}
The action of general modified gravity is
\begin{equation}
S=\int d^{4}x\sqrt{-g}\left( \frac{R+F(R,\mathcal{G},\Box R,\Box^{-1}R,...)}{2\kappa ^{2}}+\mathcal{L}_{\mathrm{m%
}}\right) \ ,  \label{Hm0}
\end{equation}%
where $\kappa^2=8\pi G$, $\mathcal{L}_{\mathrm{m}}$ is the matter
lagrangian density and the function $F(R,\mathcal{G},...)$ may
contain scalar curvature $R$, GB term $\mathcal{G}=R_{\mu \nu \gamma
\delta }R^{\mu \nu \gamma \delta }-4R_{\mu \nu }R^{\mu \nu }+R^2$
and any contributions of $\Box R$. At follows, we focus our
attention only on $F(R)$ and $F(\mathcal{G})$. By varying the action
over $g_{\mu\nu}$, the field equations can be obtained \cite{Ody1}.
The field equations corresponding to FRW equations in a spatially
flat universe with $R=6\dot{H}+12H^{2}$, in $F(R)$ gravity is
\cite{Od0910}:
\begin{eqnarray}
\rho_{\mathrm{eff}}&=&\frac{1}{\kappa ^{2}}[-\frac{F(R)}{2}+3(
H^{2}+\dot{H}) F^{\prime }(R)\nonumber\\&&-18(
4H^{2}\dot{H}+H\ddot{H}) F^{\prime
\prime }(R)]+\rho_{\mathrm{matter}},\nonumber\\
p_{\mathrm{eff}}&=&\frac{1}{\kappa ^{2}}[\frac{F(R)}{2}-(
3H^{2}+\dot{H}) F^{\prime }(R)\nonumber\\&&+6(
8H^{2}\dot{H}+6H\ddot{H}+4\dot{H}^2+\dddot{H}) F^{\prime \prime
}(R)\nonumber\\ &&+36(4H\dot{H}+\ddot{H})^2 F^{\prime
\prime\prime}(R)]+p_{\mathrm{matter}},\label{FR}
\end{eqnarray}
and in GB modified gravity, $R+F(\mathcal{G})$, with
$\mathcal{G}=24H^2(H^2+\dot{H})$, is
\begin{eqnarray}
\rho_{\mathrm{eff}}&=&\frac{1}{2\kappa
^2}[-F(\mathcal{G})+\mathcal{G} F^{\prime
}(\mathcal{G})-(24)^2H^4\nonumber\\&&\times(
4H^{2}\dot{H}+H\ddot{H}+2\dot{H}^2) F^{\prime \prime
}(\mathcal{G})]+\rho_{\mathrm{matter}},\nonumber\\
p_{\mathrm{eff}}&=&\frac{1}{2\kappa ^{2}}[F(\mathcal{G})+(24)^2H^2(
3H^{4}+20H^2\dot{H}^2+6\dot{H}^3\nonumber\\&&+4H^3\ddot{H}+H^2\dddot{H})
F^{\prime \prime}(\mathcal{G})-(24)^3H^5\nonumber\\
&&\times(2\dot{H}^2+H\ddot{H}+4H^2\dot{H})^2F^{\prime
\prime\prime}(\mathcal{G})]+p_{\mathrm{matter}}. \label{FGB}
\end{eqnarray}
Hear $\rho_{\mathrm{eff}}$ and $p_{\mathrm{eff}}$ are effective
energy density and pressure caused by extra gravitational terms due
to the modification of the GR Lagrangian. It has been showed that by
getting the effective gravitational pressure and energy density, the
equation of motion for arbitrary modified gravity can be rewritten
in the standard form of FRW in GR as \cite{Od0910}
\begin{equation}
H^2=\frac{\kappa^2}{3}\rho_{eff},\qquad
p_{eff}=-\frac{1}{\kappa^2}(2\dot{H}+3H^2).\label{SFRW}
\end{equation}
\section{A brief review on ADE model}
The metric of a general spatially flat FRW universe is given by
\begin{equation}
ds^{2}=-dt^{2}+a^{2}(t)\left[ dr^{2}+r^{2}(d\theta ^{2}+\sin
^{2}\theta d\varphi ^{2})\right],
\end{equation}
where $a(t)$ is the dimensionless scale factor. The energy density
of ADE is given by \cite{Cai1}
\begin{equation}
\rho _{D}=\frac{3n^{2}}{\kappa^{2}T^{2}},
\end{equation}%
where $n$ is ADE constant parameter and $T$ is the age of the
universe which is given by
\begin{equation}
T=\int_{0}^{t}dt=\int_{0}^{a}\frac{da}{Ha},
\end{equation}%
where $\dot{T}=1.$ The dimensionless dark energy density is defined as%
\begin{equation}
\Omega _{D}=\frac{\rho _{D}}{\rho _{cr}}=\frac{n^{2}}{T^{2}H^{2}},
\end{equation}%
where $\rho _{cr}=3H^{2}/\kappa^2$ is critical energy density. Let
us consider the dark energy dominated universe. In this case the
dark energy evolves according to its conservation law
\begin{equation}
\dot{\rho }_{D}+3H\rho_{D}(1+w_{D})=0,
\end{equation}%
where $w_{D}=p_{D}/\rho_{D}$, is equation of state of ADE, which is
\cite{Cai1}
\begin{equation}
w_{D}=-1+\frac{2\sqrt{\Omega _{D}}}{3n};~~~~~\Omega
_{D}=1.\label{wd1}
\end{equation}%
\section{Reconstruction of $F(R)$ gravity with ADE}
Hear we use ADE model to reconstruct a consistent $F(R)$. The first
FRW equation of a spatially flat universe containing an agegraphic
dark energy without any matter component can be obtained as
\cite{Cai1}
\begin{equation}
H^{2}=\frac{\kappa^2}{3}\rho_{D}=\frac{n^{2}}{T^{2}}. \label{FRWage}
\end{equation}%
By using a new variable $N=\ln \frac{a}{a_{0}}$, which is often
called e-folding, instead of the cosmological time $t$, we have
$\frac{d}{dt}=H\frac{d}{dN}$ and therefore $\frac{d^{2}}{
dt^{2}}=H^{2}\frac{d^{2}}{dN^{2}}+H\frac{dH}{dN}\frac{d}{dN}$. In a
dark energy dominated universe, without any matter component, Eq.
(\ref{FR}) can be written as
\begin{eqnarray}
&\kappa ^{2}\rho_{\mathrm{eff}}&=-\frac{F(R)}{2}+3\left(
H^{2}+HH^{\prime }\right) F^{\prime }(R)\nonumber\\&&-18\left(
4H^{3}H^{\prime}+H^2H^{\prime 2}+H^{3}H^{\prime \prime }\right)
F^{\prime \prime }(R). \label{RZ4}
\end{eqnarray}%
Here $H^{\prime }\equiv dH/dN$ and $H^{\prime \prime }\equiv
d^2H/dN^{2}$. Using $G\left( N\right) =H^{2}$, the Eq. (\ref{RZ4})
may be written as:
\begin{eqnarray}
&\kappa^{2}\rho_{\mathrm{eff}}&=-\frac{F(R)}{2}+3\left[G(N)+\frac{1}{2}G^{\prime}(N)\right]
F^{\prime}\nonumber\\&&-9G(N)\left[4G^{\prime}(N)+G^{\prime\prime}(N)\right]
F^{\prime\prime}(R), \label{RZ11}
\end{eqnarray}
and the scalar of curvature become
\begin{equation}
R=3G^{\prime }(N)+12G(N).  \label{NR}
\end{equation}
Note that from Eq. (\ref{NR}), $N$ generally is a function of $R$.
In order to reconstruct $F(R)$ gravity with ADE, by comparing Eqs.
(\ref{FRWage}) and (\ref{SFRW}), it is required to get
$\rho_{D}=\rho_{\mathrm{eff}}$. Also the age of the universe $ T$ is
a function of $N$. Using
\begin{eqnarray}
&&G(N)=\frac{n^2}{T^2(N)};\nonumber\\
&&G^{\prime}(N)=\frac{-2\sqrt{G(N)}}{T(N)}=\frac{-2n}{T^2(N)};\nonumber\\
&&G^{\prime \prime }(N)=\frac{4}{T^{2}(N)}, \label{GGp}
\end{eqnarray}
and Eq. (\ref{NR}), $T(N(R))$ can be calculated as a function of $R$
as
\begin{equation}
T(N(R))=\sqrt{\frac{6n(2n-1)}{R}}.  \label{TR}
\end{equation}%
Now from (\ref{TR}) and (\ref{GGp}), all functions $G,$ $G^{\prime
}$ and $G^{\prime \prime }$ may be rewritten as a function of $R$.
Inserting those in Eq. (\ref{RZ11}), one can obtain
\begin{equation}
2R^{2}F^{\prime \prime }(R)+(n-1)RF^{\prime}(R)-(2n-1)F(R)-nR=0.
\label{FReq}
\end{equation}
This differential equation should be solved to find a consistent
modified gravity with ADE in flat space. Its solution is
\begin{equation}
F(R)=C_{+}R^{m_{+}}+C_{-}R^{m_{-}}-R,
\end{equation}
where $m_{\pm}$ are
\begin{equation}
m_{\pm}=\frac{3-n\pm \sqrt{n^2+10n+1}}{4},\label{mpm}
\end{equation}
and $C_{\pm}$ are any arbitrary constant which are given by initial
conditions. In order to generating an accelerating expansion at the
present universe, let us consider that f(R) could be a small
constant at present universe, which is,
\begin{equation}
F(R_{0})=-3R_{0},\qquad \underset{R\rightarrow 0}{\lim
}F(R)=0,\qquad F^{\prime} (R_{0})\sim 0.
\end{equation}
Hear $R_{0}$ is current curvature $R_{0}\sim (10^{-33}eV)^2$
\cite{Od0707.1941}. Therefore constants $C_{\pm}$ are
\begin{equation}
C_{\pm}=\mp\frac{R_{0}(n-5\pm
\sqrt{n^2+10n+1})}{R_{0}^{m_{\pm}}\sqrt{n^2+10n+1}}\label{cpm}.
\end{equation}
As we see from Eqs (\ref{mpm}) and (\ref{cpm}), by choosing
$n^2+10n+1\geq 0$, a consistent $F(R)$ can be found. Therefore we
can obtain two following conditions
\begin{equation}
n\geq -0.1~~~or~~~n\leq -9.9 \label{condn}.
\end{equation}
By getting $w_{\mathrm{eff}}=w_{D}$, and from Eqs. (\ref{wd1}) and
(\ref{SFRW}), the effective EOS may be obtained as:
$w_{\mathrm{eff}}=-1+2/3n$. Hence, the constant ADE parameter as a
function of $w_{\mathrm{eff}}$ can be written as:
$n=2/3(1+w_{\mathrm{eff}})$. Therefore the conditions (\ref{condn})
may be rewritten as
\begin{eqnarray}
&&9w_{\mathrm{eff}}^2+78w_{\mathrm{eff}}+73\geq 0;
\nonumber\\
&& w_{\mathrm{eff}}\geq -1.067~~~
or~~~w_{\mathrm{eff}}\leq-7.6.\label{cpmwd}
\end{eqnarray}
In this case a transition between deceleration
($w_{\mathrm{eff}}>-1/3$)-acceleration ($w_{\mathrm{eff}}<-1/3$)
phase of the universe has been permitted and it is possible that the
dark energy dominated universe may live at effective phantom era
($w_{\mathrm{eff}}<-1;\quad n<0$). As we see from (\ref{cpmwd}), a
quintessence era, where ($w_{\mathrm{eff}}>-1;\quad n>0$) can also
be permitted in forward reconstruction method.

It is worthwhile to mention that the differential equation
(\ref{FReq}) can also be obtained from another way, followed by
Refs. \cite{Set1,Set2}. By given a quintessence scale factor form
as: $a=a_{0}t^h$ with $h>0$, or by properly shifting of time,
Phantom scale factor form: $a=a_{0}(t_{s}-t)^h$ with $h<0$, which
tell us that there will be a Big Rip singularity at $t=t_{s}$
\cite{Od0506212}. Using latter form of scale factor, we can easily
find
\begin{eqnarray}
&&T(t)=(t_{s}-t);\qquad H(t)=\frac{-h}{(t_{s}-t)};\nonumber\\&&
R(t)=\frac{12h^2-6h}{(t_{s}-t)^2} \label{ts1},
\end{eqnarray}
where $h$ is an arbitrary negative constant. Using Eq.
(\ref{FRWage}), we see that $h=n$. From Eq. (\ref{ts1}), we have
$(t_{s}-t)^2=6n(2n-1)/R$ and
$\rho_{D}=\rho_{\mathrm{eff}}=nR/(2\kappa^2(2n-1))$ and finally Eq.
(\ref{FR}) for $\rho_{\mathrm{m}}=0$, is exactly similar to Eq.
(\ref{FReq}) which is obtained in the forward way. We must mention
that by this method, the reconstruction is permitted only in phantom
era ($h<0$), or in quintessence era ($h>0$), according to choose
each form of mentioned scale factor, separately.
\section{Reconstruction of $F(\mathcal{G})$ gravity with ADE}
In $F(\mathcal{G})$ gravity, like $F(R)$ gravity, by using the
variable $N$ instead of the cosmological time $t$, Eq. (\ref{FGB}),
without any matter component, may be rewritten as
\begin{eqnarray}
&\kappa
^{2}\rho_{\mathrm{eff}}&=-\frac{F(\mathcal{G})}{2}+12H^2\left(
H^{2}+HH^{\prime }\right) F^{\prime
}(\mathcal{G})\label{GB4}\\&&-(12)^2H^6\left( 6H^{\prime
2}+8HH^{\prime}+2HH^{\prime \prime }\right) F^{\prime \prime
}(\mathcal{G}).\nonumber
\end{eqnarray}%
Using $G\left( N\right) =H^{2}$, the GB term is
\begin{equation}
\mathcal{G}=12G(N)\left[2G(N)+G^{\prime }(N)\right]\label{GBt},
\end{equation}
and the Eq. (\ref{GB4}) may be written as
\begin{eqnarray}
\kappa^{2}\rho_{\mathrm{eff}}&=&-\frac{F(\mathcal{G})}{2}+6\left[2G^{2}(N)+G(N)G^{\prime
}(N)\right] F^{\prime }(\mathcal{G})\nonumber\\
&&-(12)^2G(N)^2[G^{\prime
2}(N)+4G(N)G^{\prime}(N)\nonumber\\&&+G(N)G^{\prime \prime }(N)]
F^{\prime \prime }(\mathcal{G}). \label{GB5}
\end{eqnarray}
Note that from Eq. (\ref{GBt}), $N$ generally is a function of
$\mathcal{G}$. In order to reconstruct $F(\mathcal{G})$ gravity with
ADE, it is required to get $\rho_{D}=\rho_{\mathrm{eff}}$. Also from
Eqs. (\ref{GBt}) and (\ref{GGp}), $T(N(\mathcal{G}))$ can be
calculated as a function of $\mathcal{G}$ as
\begin{equation}
T(N(\mathcal{G}))=\left(\frac{24n^3(n-1)}{\mathcal{G}}\right)^{\frac{1}{4}}.
\label{TG}
\end{equation}%
Now from (\ref{TG}) and (\ref{GGp}), all functions $G$, $G^{\prime
}$ and $G^{\prime \prime }$ can be calculated as a function of
$\mathcal{G}$. Inserting those into Eq. (\ref{GB5}), we obtain
\begin{eqnarray}
&&8\mathcal{G}^{2}F^{\prime \prime
}(\mathcal{G})+2(n-1)\mathcal{G}F^{\prime}(\mathcal{G})-2(n-1)F(\mathcal{G})\nonumber\\&&-\sqrt{6n(n-1)\mathcal{G}}=0.
\label{FGBeq}
\end{eqnarray}
Its solution can be obtained as
\begin{equation}
F(\mathcal{G})=C_{1}\mathcal{G}+C_{2}\mathcal{G}^{(-\frac{n-1}{4})}-\frac{\sqrt{6n(n-1)}}{n+1}\sqrt{\mathcal{G}},
\end{equation}
and as a function of $w_{\mathrm{eff}}$, it is rewritten as
\begin{equation}
F(\mathcal{G})=C_{1}\mathcal{G}+C_{2}\mathcal{G}^{(\frac{1}{12}\frac{1+3w_{\mathrm{eff}}}{1+w_{\mathrm{eff}}})}-%
\frac{\sqrt{-12(1+3w_{\mathrm{eff}})}}{5+3w_{\mathrm{eff}}}\sqrt{\mathcal{G}}.
\end{equation}
We see that a consistent $R+F(\mathcal{G})$ gravity may be existed,
provided that $w_{\mathrm{eff}}\leq -1/3$.
\section{conclusion}
In this paper we show that a consistent modified $F(R)$ and
$F(\mathcal{G})$ gravities may be reconstructed forwardly so that it
gives the cosmological evolution of ADE model in a no matter
spatially flat universe with no need of the hand insertion of extra
dark components. After calculating a consistent $F(R)$ with ADE, we
obtain conditions for $w_{\mathrm{eff}}$ and see that reconstruction
is possible for both phantom and non-phantom era. These calculations
have also been done for $F(\mathcal{G})$ gravity and the condition
for a consistent $F(\mathcal{G})$ is obtained. Although it is
possible that dark energy dominated universe live (or enter) at
effective phantom era like non-phantom era, deceleration phase of
the universe ($w_{\mathrm{eff}}>-1/3$) is not achieved in this case.


\begin{thebibliography}{99}

\bibitem{Hann} S. Hannestad and E. Mortsell, Phys. Rev. D 66, 063508
(2002); A. Melchiori, L. Mersini-Houghton, C. J. Odman, and M.
Trodden, Phys. Rev. D 68, 043509 (2003);H. Jassal, J.Bagla, and T.
Padmanabhan [astro-ph/0506748].

\bibitem{Stra} A. A. Starobinsky, Phys. Lett. B \textbf{91}, 99 (1980).

\bibitem{Ker}R. Kerner, Gen. Rel. Gravit. \textbf{14}, 453 (1982); J. P.
Duruisseau, and R. Kerner, Class. Quantum Grav. \textbf{3}, 817
(1986).

\bibitem{Noj1}
S.~Nojiri and S.~D.~Odintsov, J. Phys. Conf. Ser. {\bf 66}, 012005
(2007).

\bibitem{Noj2}
S.~Nojiri and S.~D.~Odintsov, Phys. Rev. D {\bf 74}, 086005 (2006);
   J.\ Phys.\ A  {\bf 40}, 6725 (2007); S.~Capozziello, S.~Nojiri,
    S.~D.~Odintsov and A.~Troisi, Phys. Lett. B {\bf 639}, 135 (2006);
    E.~Elizalde and D.~Saez-Gomez,
    [arXiv:0903.2732].

\bibitem{Cruz}
A.~de la Cruz-Dombriz and A.~Dobado, Phys. Rev. D {\bf 74}, 087501
(2006); J.~L.~Cortes and J.~Indurain, Astropart. Phys. {\bf 31}, 177
(2009); I.~H.~Brevik, Gen. Rel. Grav. {\bf 38}, 1317 (2006);
L.~N.~Granda, [arXiv:0812.1596]; X.~Wu and Z.~H.~Zhu, Phys. Lett. B
{\bf 660}, 293 (2008).

\bibitem{Bam}
K.~Bamba, C.~Q.~Geng, S.~Nojiri and S.~D.~Odintsov, Phys. Rev. D
{\bf 79}, 083014 (2009); K.~Bamba, S.~Nojiri and S.~D.~Odintsov,
JCAP {\bf 0810}, 045 (2008); K.~Bamba and C.~Q.~Geng,
[arXiv:0901.1509].

\bibitem{Noj3} S.~Nojiri and S.~D.~Odintsov, and D.
S\'{a}ez-G\'{o}mez, [arxiv:0908.1209].

\bibitem{Eliz} E. Elizalde, R. Myrzakulov, V.V. Obukhov and D.
S\'{a}ez-G\'{o}mez, [arxiv:10013636].

\bibitem{Set1} M.~R.~Setare, Int. J. Mod. Phys. D {\bf 17}, 2219 (2008).

\bibitem{Khod1} H. Wei and R. G. Cai [arxive:07074526]; A. Khodam-Mohammadi. and M. Malekjani, [arxiv:];
M. Malekjani and A. Khodam-Mohammadi, [arxiv:].

\bibitem{Cai1} R. G. Cai,  Phys. Lett. B \textbf{657}, 228 (2007).

\bibitem{Maz} M. Maziashvili,  Int. J. Mod. Phys. D \textbf{16}, 1531 (2007);
Phys. Lett. B \textbf{652}, 165 (2007).

\bibitem{Cai3} H. Wei, and R.G. Cai, Phys. Rev. D \textbf{71}, 043504 (2005);
H. Wei, and S.N. Zhang, Phys. Lett. B \textbf{644}, 7(2007); A.
Sheykhi, Phys. Rev. D \textbf{81}, 023525 (2010).

\bibitem{Ody1} S.~Nojiri and S.~D.~Odintsov, eConf {\bf C0602061}, 06
(2006), [Int.\ J.\ Geom.\ Meth.\ Mod.\ Phys.\  {\bf 4}, 115 (2007)].

\bibitem{Od0910} S.~Nojiri and S.~D.~Odintsov, [arxive:09101464][hep-th].

\bibitem{Cai0707} R. G. Cai, [arxiv:0707.4049][hep-th].

\bibitem{Od0506212} S.~Nojiri and S.~D.~Odintsov, [hep-th/0506212].

\bibitem{Set2} M.R. Setare, [arxiv:0908.0196][gr-qc].

\bibitem{Karami1} Karami, K., Sorouri, A.: Phys. Scr. 82, 025901
(2010).

\bibitem{Malek2} Malekjani, M., Khodam-Mohammadi, A.:
[arXiv:1004.1017v2](2010).

\bibitem{Od0707.1941} S.~Nojiri and S.~D.~Odintsov, Phys. Lett. B {\bf 657}, 238
(2007).

\end{thebibliography}
\end{document}